\documentclass[12pt]{article}

 \usepackage{epsfig}
 \usepackage{amssymb}

 \def\tb{\ensuremath{\tan\beta}}

 \def\M{\mathcal{M}} \def\MSQ{\overline{|\M|^2}}

  \def\ckm{\ensuremath{V_{\rm CKM}^2}}
 \def\aem{\ensuremath{\alpha_{\rm EM}}} \def\as{\ensuremath{\alpha_s}}
  
  \def\nc{\ensuremath{N_C}}
     
 \def\slash{/\kern -5pt}
 \def\etmiss{\ensuremath{\kern 3pt\big{/}\kern -9pt E_T}}
  
  \def\ims #1 {\ensuremath{M^2_{[#1]}}}
 \def\sm{\ifmmode{{\rm SM}}\else{Standard Model}\fi}
 \def\qcd{\textsc{Qcd}} \def\MC{Monte Carlo} 
 \def\susy{\ifmmode{\rm SUSY}\else{supersymmetric}\fi}  
 \def\mssm{\ifmmode{\rm MSSM}\else{\textsc{Mssm}}\fi}
 \def\Susy{Supersymmetric} \def\MSSM{Minimal \Susy\ \sm}
  \def\vegas{\texttt{VEGAS}}
 \def\form{\texttt{FORM}}

 \def\pl #1 #2 #3 {{\it Phys.\ Lett.} {\bf#1} (#2) #3}
 \def\np #1 #2 #3 {{\it Nucl.\ Phys.} {\bf#1} (#2) #3}
 \def\zp #1 #2 #3 {{\it Z.\ Phys.} {\bf#1} (#2) #3}
 \def\pr #1 #2 #3 {{\it Phys.\ Rev.} {\bf#1} (#2) #3}
 \def\prep #1 #2 #3 {{\it Phys.\ Rep.} {\bf#1} (#2) #3}
 \def\prl #1 #2 #3 {{\it Phys.\ Rev.\ Lett.} {\bf#1} (#2) #3}
 \def\mpl #1 #2 #3 {{\it Mod.\ Phys.\ Lett.} {\bf#1} (#2) #3}
 \def\rmp #1 #2 #3 {{\it Rev.\ Mod.\ Phys.} {\bf#1} (#2) #3}
 \def\jp #1 #2 #3 {{\it J.\ Phys.} {\bf#1} (#2) #3}     
 \def\cpc #1 #2 #3 {{\it Comp.\ Phys.\ Comm.} {\bf#1} (#2) #3}
 \def\epj #1 #2 #3 {{\it Eur.\ Phys.\ J.} {\bf#1} (#2) #3}
 \def\jhep #1 #2 #3 {{\it JHEP} {\bf #1} (#2) #3}
 \def\ibidem #1 #2 #3 {{\it ibidem} {\bf#1} (#2) #3}
 \def\xx #1 #2 #3 {{\bf#1}, (#2) #3}
 \def\preprint{{\it preprint}}
 \def\cavendish #1 {\preprint\ Cavendish--HEP--#1}

\begin{document}
\thispagestyle{empty}
\setcounter{page}{0}

\begin{flushright}
{RAL--TR--1999--012}\\
{January 1999\hspace*{.5 cm}}\\
\end{flushright}

\vspace*{\fill}

\begin{center}
{\Large \bf 
Searching for heavy charged Higgs bosons\\[0.25cm]
in the neutrino--tau decay mode at LHC}\\[1.5cm]
{\large Kosuke Odagiri}\\[0.4cm]
{\it Rutherford Appleton Laboratory,}\\
{\it Chilton, Didcot, Oxon OX11 0QX, UK.}\\[0.5cm]
\end{center}
\vspace*{\fill}

 \begin{abstract}
 {\noindent 
 We discuss the search for the heavy charged Higgs bosons $H^\pm$,
implicitly of the \MSSM, in the $\tau\nu$ decay mode at the LHC. Compared
to the dominant decay mode $H^\pm\to bt$, the channel suffers from
suppression due to the branching ratio and the lack of direct mass
reconstruction, but the reduced \qcd\ background makes it a feasible
channel especially in the large \tb\ region. We study the production in
`$bt$ fusion' via $bg\to tH^-$, and the leading irreducible background
$bg\to tW^-$.
 Our results indicate that for the $H^\pm$ mass of greater than 200 GeV
and up to 1 TeV and higher, they can be discovered in this channel for a
vast range of the parameter space, down to at least $\tan\beta\sim3$ and
potentially the whole range of $\tan\beta$ down to 1.5 if the signal
selection efficiency could be improved fourfold.
 Our analysis is sensitive to top quark identification at large rapidity,
and should be supplemented with a full study including jet showering and
detector effects.
 }
 \end{abstract}

\vspace*{\fill}
\newpage

\section{Introduction}

 The charged Higgs bosons $H^\pm$ are a central prediction of the Two
Higgs Doublet Model which is necessitated by supersymmetry.
 Given the results of recent experimental analyses, the neutral Higgs
exclusion at LEP2 \cite{lep2search} and the constraint on the rare bottom
quark decay $b\to s\gamma$ \cite{bdecay}, both of which favour the elusive
heavy mass region for $H^\pm$, at least in the \mssm, it is important to
establish the parameter regions in which $H^\pm$ can be discovered at
present and future colliders.

 It is believed \cite{mo_lep2} that LEP2 can not discover the charged
Higgs bosons in the mass regions indicated by the \MSSM\ (\mssm) and the
current bound \cite{lep2search} on neutral Higgs bosons, viz the mass
relation $M^2_{H^\pm}=M^2_W+M^2_{A^0}\gtrsim110$ GeV. The search for such
a particle needs greater centre-of-mass energy, notably at LHC and, to a
more limited extent, Tevatron \cite{spira}. The purpose of this paper is
to establish the significance of the secondary $H^\pm$ decay channels,
mainly at LHC.

 If $H^\pm$ are `heavy', which is taken from here on to imply
$M_{H^\pm}\gtrsim m_t$, they can not be produced in top decay and the main
production mode for them at hadron colliders is `$bt$ fusion', viz:
 \begin{equation}\label{gbth} gb\to tH^-. \end{equation}
 We have taken $b$ as a sea quark, as is appropriate at such high energy
collisions, whereas $t$ is created via $g\to t{\bar t}^*$. Of course the
charge conjugate process $g\bar b\to\bar tH^+$ must also be included, and
this behaves in exactly the same way since there is no intrinsic $b$
component of the proton.

 The above production channel has been studied extensively in the
literature \cite{heavysearches}. Recent analyses have focussed on the
dominant decay mode $H^+\to t\bar b$ and the reduction of the \qcd\
background associated with it. The utilisation of the less dominant decay
modes was discussed in the past \cite{hunters}. Here we focus on the decay
mode $H^+\to\tau^+\nu_\tau$, improving on previous analyses by adopting
the latest numbers for \sm\ parameters and parton distribution functions
as well as using kinematic selection techniques and the tau polarisation
analysis \cite{bkm}.

 The result of our analysis indicates that the discovery region for
$H^\pm$ is far greater than has previously been thought, covering the
whole of the $(M_{H^\pm}, \tan\beta)$ parameter space up to
$M_{H^\pm}\sim1$ TeV and higher, and down to at least $\tan\beta\sim3$. 
Optimisation in the analysis procedure and parameters can potentially
extend the lower limit in $\tan\beta$ to 1.5, below which is excluded by
LEP2 measurements so far in the \mssm.

\section{Calculation}

 In order to minimise the parameter dependence of our calculations and to
make them applicable to as wide a range of parameters as possible, we
adopt $M_{H^\pm}$ and $\tan\beta$ as the two Higgs sector parameters. When
considering the $H^\pm$ decays, let us furthermore assume that the neutral
Higgs mixing angle $\alpha$ is given by $\beta-\alpha=\pi/2$ which holds
in \mssm\ at tree level in the heavy mass limit of the charged Higgs
bosons, explicitly, $M^2_{H^\pm}\gg M^2_Z$. In this limit, practically the
only \sm\ decay modes are $tb$ and $\tau\nu$, and we assume superpartners
too heavy to be produced in $H^\pm$ decays.

 We restrict our discussions to the tree level, since our primary purpose
is to establish the plausibility of signal detection, rather than to
propose precision measurements. We adopt MRS 1998 leading order parton
distributions \cite{mrs98}, and calculate the strong coupling at one loop
using the default MRS parameter values, neglecting virtual top
contributions in both cases.

 We assume zero kinematic bottom quark mass to be consistent with the sea
quark picture we are adopting, whereas we retain finite bottom quark pole
mass $m_b$ for the Yukawa coupling, which we take to be 4.25 GeV.
Similarly, we take the tau kinematic mass to be zero and set the Yukawa
coupling mass to 1.78 GeV. Both kinematic and Yukawa masses of the top
quark are taken to be 175 GeV. The top quark, $H^\pm$ and $W^\pm$ widths
are calculated at tree level, although we adopt the narrow width
approximation for the top quark and $H^\pm$. We also note that the quark
masses implicit in the parton distribution functions are independent
parameters which are in general distinct from the kinematic and Yukawa
masses we choose to adopt.

 The electroweak parameters are $\aem=1/128$, $\sin^2\theta_W=0.2315$,
$M_Z=91.187$ GeV, $M_W=M_Z\cos\theta_W\approx79.94$ GeV.
 The Cabbibo--Kobayashi--Maskawa matrix element $\ckm[bt]$ is taken to be
1 (it is practically 1. See \cite{pdg}).

 As for the LHC collider parameters \cite{lhc}, the centre-of-mass energy
for $pp$ collision is 14 TeV. Integrated luminosities of 10 fb$^{-1}$ and
100 fb$^{-1}$ per year are expected in the low luminosity and high
luminosity options, respectively.

 The squared matrix elements are calculated by hand with the usual trace
method, convoluted with the MRS parton distribution functions, and
integrated using \vegas\ \cite{vegas}. The results are tested against the
literature \cite{hunters} where they are available, and tests of gauge
independence were carried out by hand.

 When computing decay level distributions, the top quark is assumed to be
on the mass shell using the so-called `narrow width approximation', in
order to minimise the number of `unphysical' Feynman graphs we must
consider without violating gauge invariance. We similarly take the $W^\pm$
boson arising from top quark decay to be on shell.

 For the signal events, we assume that the top quark decays independently
of production, by neglecting the spin correlation effects which are anyhow
small for our purposes, in order to maximise the parameter space that can
be covered in our analysis. Since we take top quark to be on the mass
shell, this procedure is exact in the limit of identical left and right
$H^\pm$ coupling, at $\tan\beta=\sqrt{m_t/m_b}$. We also take $H^\pm$ to
be on the mass shell since we do not consider the finite width effect to
be important at this stage.

 For the background we incorporate the exact decay distribution of the top
quark, and the finite constant width of $W^\pm$, both of which are
potentially important when considering kinematic cuts to reduce the
background. The calculation was tested using \form\ \cite{form}.

 The theoretical uncertainties associated with the above procedure are
expected to be dominated by two sources. Firstly, the \qcd\ higher order
corrections which are of the order $\as(M_{H^\pm})\sim10\%$. To be more
precise, our procedure will need to be tested for jet and detector
effects, which must be carried out using \MC\ simulations where the
leading \qcd\ corrections are resummed and factorised.
 Secondly, as we discuss in the next chapter, the possibility of the other
decay modes, such as the $H^\pm\to hW^\pm$ decay mode for relatively light
$H^\pm$ at low $\tan\beta$.


\section{Analysis strategy}

 The tree level Feynman diagrams for the signal process are shown in
figure \ref{feyn}. The spin- and colour-averaged matrix element squared
for the process simplifies as follows:
 \begin{eqnarray}\label{mgbth}\MSQ(g_{(1)}b_{(2)}\to t_{(3)}H^-_{(4)})&=&
 \left(\frac{g^2_s}{2\nc}\right)\left(\frac{e^2}{2\sin^2\theta_W}\right)
 \left(\frac{m_b^2\tan^2\beta+m_t^2\cot^2\beta}{2M_W^2}\right)
 \times\nonumber\\&&\times\left(-\frac{u^2_4}{st_3}\right)
 \left[1+2\frac{m_4^2-m_3^2}{u_4}
       \left(1+\frac{m_3^2}{t_3}+\frac{m_4^2}{u_4}\right)\right].
 \end{eqnarray}
 Here $g_s$ is the strong coupling, $g_s^2/4\pi=\as$, and as a first
approximation we evaluate it at scale $Q=(m_3+m_4)$. $N_c=3$ for \qcd. 
$s$, $t$ and $u$ are the usual Mandelstam variables, defined with the
partonic momenta. $t_3=t-m_3^2=t-m_t^2$ and $u_4=u-m_4^2=u-M^2_{H^\pm}$.
$\tan\beta$ is the ratio of the two expectation values of the two Higgs
doublets, and we assume $1<\tan\beta<m_t/m_b$.

 We do not tag the spectator bottom quark. However, in order to reduce the
background, for example, from top pair, $W^\pm Z^0$ and single $W^\pm$
production, and to have a grip on the irreducible background, we do make
use of the `spectator' top quark, $t_{(3)}$. In \cite{hunters,
stifflepton}, the following `stiff lepton' procedure was proposed: the top
quark moves nearly parallel to the original $g_{(1)}$ parton direction,
and will be lost in the beam jets.  However, if it decays leptonically,
the electron or muon will be kicked out with $p_T$ of the order $m_t/2$ so
that this can be used to trigger the signal events. Potentially the bottom
quark can also be tagged, especially when it decays leptonically.

 In this study we consider both the leptonic and hadronic decays. Although
the leptonic mode is more inclusive, the hadronic branching fraction is
larger ($BR[{\rm hadronic}]=2/3$), and the lack of the missing momentum
from leptonic decay makes the handling of the kinematic distributions more
transparent.

 In the leptonic decay mode, we consider the inclusive sample of top-like
events, wherein we assume the top quark identification to be carried out
with a universal efficiency of $\varepsilon_t$, with $\varepsilon_t\sim
BR[{\rm leptonic}] = 2/9$.

 In the hadronic decay mode, we require the inclusive multi-jet final
state to have a reconstructed mass near the top mass. If bottom jet 
tagging can be utilised, this will aid the procedure by allowing the
$W^\pm$ mass to be reconstructed also. We can not make a full simulation
of the top quark identification efficiency, but as a first approximation
we require all parton level decay products to be reasonably away from the
beam directions, at pseudorapidities of less than 1.5 in order to reduce
the \qcd\ noise. We do not impose cuts on separation between top decay
products.

 It has been shown \cite{stifflepton}, and we will assume, that the
tagging of the top quark is very effective in reducing the background,
eliminating all but the irreducible background:
 \begin{equation}\label{gbtw} gb\to tW^-, \end{equation}
 whose presence is the main obstacle to the observation of the signal
(\ref{gbth}).

 Let us assume the leptonic decay mode $H^-\to\tau\bar\nu$. In the limit
of heavy $H^\pm$, $M^2_{H^\pm}\gg m^2_t$, the branching ratio is easily
seen to be given by:
 \begin{equation}\label{bratios}
 BR[\tau\nu]\approx\frac{\Gamma[\tau\nu]}{\Gamma[bt]}=
\frac{m_\tau^2\tan^2\beta}{\nc(m_t^2\cot^2\beta+m_b^2\tan^2\beta)}.
 \end{equation}
 For finite $M_{H^\pm}$ the branching ratio is always greater than this
limit, as there is kinematic suppression on top quark for the $H^\pm\to
bt$ decay mode, which goes as $(1-m_t^2/M^2_{H^\pm})$. Thus if there are
no other decay modes:
 \begin{equation}\label{exactbratios}
 BR[\tau\nu]=\left[1+\frac{m^2_b\nc}{m^2_\tau}
             \left(1+\frac{m^2_t}{m^2_b\tan^4\beta}\right)
             \left(1-\frac{m^2_t}{M^2_{H^\pm}}\right) \right]^{-1}.
 \end{equation}

 Figure \ref{brplot}a shows the $M^2_{H^\pm}\gg m^2_t$ limit of equation
(\ref{exactbratios}), as a function of $\tan\beta$. By requiring the other
decay modes to vanish, we have implicitly assumed $\cos^2(\beta-\alpha)=0$
which holds in the heavy mass limit in the \mssm\ at tree level. The
$H^\pm\to W^\pm h^0$ decay rate is proportional to this. The ratio of this
mode to the $tb$ mode is given by:
 \begin{equation}\label{whmode}
 \frac{\Gamma[W^\pm h^0]}{\Gamma[bt]}=
 \frac{\lambda^{3/2}[M_{H^\pm},M_{W^\pm},M_{h^0}]\cos^2(\beta-\alpha)}
 {2\nc\lambda[M_{H^\pm},m_t,0](m_b^2\tan^2\beta+m_t^2\cot^2\beta)}
 \end{equation}
 where $\lambda[m_1,m_2,m_3]=(m_1^2-(m_2+m_3)^2)(m_1^2-(m_2-m_3)^2)$ and
is symmetric under the permutations of the three masses. At tree level in
\mssm, it can be shown that $\cos^2(\beta-\alpha)$ is given in terms of
the pseudoscalar Higgs mass $M_A$ and $\tan\beta$ by:
 \begin{equation}
 \cos^2(\beta-\alpha)=\frac{1}{2}\left[1-\left[1+
 \left(\frac{M_Z^2\sin4\beta}{M_A^2-M_Z^2\cos4\beta}\right)^2
 \right]^{-1/2}\right].
 \end{equation}
 Combined with equation (\ref{whmode}), the approximation is valid if
$M^2_A\gg M^2_Z$ which is implicit in $M^2_{H^\pm}\gg m^2_t$ since
$M_{H^\pm}>M_A$ in \mssm. However, if $m_t(+m_b)>M_{W^\pm}+M_{h^0}$, the
$W^\pm h^0$ decay could dominate in a small window, noting $M_{h^0}\gtrsim
80$ GeV, just below the threshold for the $bt$ mode.

 The factor $(m_t^2\cot^2\beta+m_b^2\tan^2\beta)$ cancels between
(\ref{mgbth}) and (\ref{bratios}), and we see that the cross section
relevant to us goes as:
 \begin{equation}\label{brnorm}
 \MSQ(gb\to tH^-\to t\tau\bar\nu_\tau)\propto
 \frac{m_\tau^2\tan^2\beta}{2\nc M_W^2}.
 \end{equation}
 In figure \ref{brplot}(b) we plot this quantity, which represents the
overall $\tan\beta$ dependence of production and decay in the $\tau\nu$
mode. The $1/BR[\tau\nu]$ contribution, neglected in (\ref{bratios}) and
(\ref{brnorm}), is also included in this plot. In addition, in our
forthcoming simulations in the next chapter, we also include the kinematic
$(1-m_t^2/M^2_{H^\pm})$ suppression on the $bt$ decay mode.

 The complete matrix element squared for the background process
(\ref{gbtw}), including the $t$ and $W^\pm$ decay, is available in
\cite{ellisparke}.  However, in order to gain insight into the structure
of the background process let us present the spin- and colour- averaged
matrix element squared for (\ref{gbtw}). The Feynman graphs are identical
to figure \ref{feyn} with the substitution $H^-\to W^-$, in the following
simple form\footnote{In our decay level simulations we supplement
(\ref{mgbtw}) by including the top quark and $W^\pm$ decay, with the top
quark being on shell, as well as the $W^\pm$ from top quark decay. For the
signal, (\ref{mgbth}) is supplemented similarly, but we set the $H^\pm$ on
shell, and assume the top quark to decay independently of production by
averaging over the top quark helicities, as mentioned in the previous
chapter. Neither of these approximations are expected to have a
significant effect on our results.}:
 \begin{eqnarray}\label{mgbtw}\MSQ(g_{(1)}b_{(2)}&\to& t_{(3)}W^-_{(4)})=
 \left(\frac{g^2_s}{2\nc}\right)\left(\frac{e^2}{2\sin^2\theta_W}\right)
 \times\nonumber\\&&\times\left[2+\left(1+\frac{m_3^2}{2m_4^2}\right)
  \left(-\frac{u^2_4}{st_3}\right) \left[1+2\frac{m_4^2-m_3^2}{u_4}
  \left(1+\frac{m_3^2}{t_3}+\frac{m_4^2}{u_4}\right)\right]\right].
 \end{eqnarray}
 This contains an `isotropic' part, the 2 inside the first square
brackets, and a `scalar' part whose kinematics are exactly proportional to
the signal cross section (\ref{mgbth}) in the limit of degenerate $H^\pm$
and $W^\pm$ masses. The ratio of the two is around 1:1 (see forthcoming
table \ref{tablea}).  As for the $W^\pm$ branching ratio into
$\tau\nu_\tau$, this equals $1/9\approx11\%$. Thus in this limit, the
ratio of signal to non-isotropic, `scalar' background, assuming $H^\pm$
branching ratio of equation (\ref{bratios}), is:
 \begin{equation}
 \frac{\sigma(gb\to tH^-\to t\tau\nu_\tau)}
      {\sigma(gb\to tW^-\to t\tau\nu_\tau)_{\rm scalar}}
 = \frac{9}{2\nc}\left(\frac{m_\tau\tan\beta}{M_W}\right)^2
   \frac{2M_W^2}{m_t^2+2M_W^2}.
 \end{equation}
 $M_W/m_\tau\sim40$, so this amounts to some excess of signal over
background for large values of $\tan\beta$. In reality, this will be
reduced further by the phase space suppression on $H^\pm$ production, but
to compensate for this, the mass difference between $H^\pm$ and $W^\pm$
can be exploited in kinematic cuts. A more subtle point is that $H^\pm$
are scalar whereas $W^\pm$ are vector, leading to different distributions
of the decay products which may be utilised once a significant excess of
signal over background is observed through the kinematic cuts. As a final
stage of the signal--background analysis, we can use tau polarisation as
described in \cite{bkm}, noting that their helicities are predominantly
left in $W^\pm$ decay, whereas they are predominantly right in $H^\pm$
decay.

 To summarise, our strategy is as follows.
       Firstly, we select events with top quarks,
       possibly identified using the stiff lepton trigger, of
       the forms $t+\tau+\etmiss+X$ in the hadronic top quark decay mode,
       or $\ell+\tau+\etmiss+X$ in the leptonic decay mode, where $\ell$
       is $e$ or $\mu$, \etmiss\ is the typically large missing transverse
       energy, and $X$ contains jets near the beam directions.
       Secondly, we utilise kinematic cuts to
       screen out $W^\pm$ events. At this stage, it is desirable to be
       able to reconstruct the $H^\pm$ mass from the transverse momentum
       distribution of tau. In experimental analyses, the kinematic
       selection cuts will be replaced by the fitting of the distributions
       with signal and background expectations, which is a more elegant
       and effective procedure.
       Thirdly, we apply tau polarisation and spin correlation analyses
       in order to confirm the presence of $H^\pm$.

\section{Results}

 The total signal cross section, corresponding to (\ref{gbth}) but
excluding the $\tan\beta$ dependence of equation (\ref{brnorm}), the
kinematic suppression on $tb$ mode of equation (\ref{bratios}), and
excluding the $bg$ initial state and the charge conjugate processes, is
shown in figure \ref{figa}.  It can be seen that the cross section falls
approximately exponentially with increasing $M_{H^\pm}$, but even for
large masses the cross section is significant, bearing in mind the
integrated luminosity of 100 fb$^{-1}$ per year in the high luminosity
option at LHC.

 In table \ref{tablea} we present the typical signal and background cross
sections including the $\tan\beta$ dependence, the kinematic suppression
on $bt$ mode, factor 4 for the $bg$ initial state and charge conjugation,
and the $W^\pm\to\tau\nu$ branching ratio of 1/9. It is clear that the
signal suffers from large background, against which we must find powerful
cuts or distributions in order to make the signal visible.

 In order to proceed further, we need to examine the distributions of
decay products. In figures \ref{figb} and \ref{figc} we present the
rapidity and transverse momentum distributions. Since all particles in our
final state are treated as being massless, the rapidity can be replaced by
pseudorapidity. We have plotted the leptonic top decay modes only. For the
hadronic top decay, discarding the small effects of spin correlation, the
light quark jet distributions are as the lepton distributions, and the
missing momentum distribution is identical to the tau transverse momentum
distribution.

 We have also computed the rapidity difference between tau and the lepton,
and found that the distribution does not behave significantly differently
from the rapidity distribution of the lepton alone.

 At first sight there seems little hope for signal detection. The rapidity
distributions are similar, the small difference being partly due to spin
correlation and partly due to the different resonance mass. The spin
correlation can not be used to reduce the background since we have made
the assumption that the top quark decays independently of production when
generating the signal plots. Anyhow, there will be a small spin
correlation effect in kinematic distributions involving the top quark
decay, which can be used to confirm the coupling structure of signal and
background events if we acquire a sufficient sample.

 The tail of transverse momentum distribution for the background does not
fall sufficiently to allow us to extract a convincing signal, especially
if $\tan\beta$ is not high enough.  This is as remarked in \cite{hunters,
stifflepton}.

 On a closer examination of the the tail of tau transverse momenta for
the background events, we note that this is not primarily due to off-shell
$W^\pm$ production, but due to $W^\pm$ being produced back-to-back with
the top quark at large transverse momenta. Indeed, the author has made a
similar plot using the `narrow width approximation' for the $W^\pm$ and
the difference between this and figure \ref{figc}c is small. At a more
analytical level too, it is easy to see that the off-shell effect is
smaller than the kinematic effect of back-to-back production.

 As noted earlier, there is an `isotropic' component in the background
production cross section (\ref{mgbtw}) which makes the background even
more prone to this back-to-back production than the signal. Having said
that, we see that at large transverse momenta, the signal distributions of
figures \ref{figc}a, b also exhibit `dips' in missing transverse momenta
at high energies corresponding to the back-to-back production.

 Such events are simpler to deal with than the `down-the-beam-pipe' type
events, since in this case the hadronic top quark decay can be used to
reconstruct the top quark mass, giving us more control over the signal
distributions.


 Let us proceed as follows.
 Firstly, we remove events where tau and missing transverse momenta are
not back-to-back, viz $\phi_{\tau-{\rm miss}}<\phi^{\rm cut}=\pi/2$ where
$\phi$ is the azimuthal angle.
 Secondly, out of the events which survive the above requirement, we
remove events where, in the leptonic case, a top quark remnant, either the
bottom quark or the lepton, is harder than either the tau or the missing
transverse momentum.  In the hadronic case, we can proceed similarly,
imposing some requirement on the reconstructed top quark (or a remnant)
transverse momentum against either the tau or the missing transverse
momentum. As default, we adopt $p_T(t)$ as the momentum cut-off.
 Thirdly, we make cuts on the tau and the missing transverse momenta,
either by an explicit cut or by a plot of the cross section against
$p_T^{\rm cut}$.

 As for the detector cuts, for the leptonic decay mode of the top quark,
the only requirement is on the tau and the missing transverse momenta.
This will also help eliminate the reducible background, for example, from
top pair production. We tentatively set this at 50 GeV. The final
transverse momentum cut mentioned above will alter this minimum value at a
later stage in the analysis. We also require the tau rapidity to be
between $-2.5$ and $+2.5$, although this may be redundant after the
transverse momentum cuts.

 For the hadronic decay mode of the top quark we require all top decay
products to have rapidity between $-1.5$ and $+1.5$, in addition to the
transverse momentum requirement above.

 To summarise, the cuts are:
 \begin{eqnarray}\label{leptoniccuts}
 \phi_{\tau-{\rm miss}}&>&\phi^{\rm cut}=\pi/2; \\
 \min(p_T(\tau),\etmiss) &>& \max(50\ {\rm GeV},p_T(b),p_T(\ell)); \\
 |\eta(\tau)| &<& 2.5
 \end{eqnarray}
 for the leptonic decay of the top quark, and
 \begin{eqnarray}\label{hadroniccuts}
 \phi_{\tau-{\rm miss}}&>&\phi^{\rm cut}=\pi/2; \\
 \min(p_T(\tau),\etmiss) &>& \max(50\ {\rm GeV},p_T(t)); \\
 |\eta(\tau)| &<& 2.5; \\
 \max(|\eta(b)|,|\eta(q)|,|\eta(q')|)&<&1.5
 \end{eqnarray}
 for the hadronic decay of the top quark.

 The resulting $p_T(\tau)$ and \etmiss\ distributions are used to
determine the optimum transverse momentum cut procedure.

 The result is shown in tables \ref{tableb} and \ref{tablec}.

 The cuts are not optimised, and it is possible that the signal against
background ratio can be improved without reducing the absolute rate too
much. In any case, it is likely that the hadronic decay of the top quark
offers a far greater chance of signal resolution, even though there is a
significant loss of top remnants near the beam pipe. The improved
background reduction is because of the absence of the neutrino from the
decay of the top quark, allowing us to know the transverse momentum of the
tau neutrino. Let us concentrate on the hadronic decay from here on. 

 In figure \ref{figd} we show the tau transverse momentum distribution for
the background events, after cuts, before the final transverse momentum
cut. The missing transverse momentum \etmiss, which we do not plot here,
should have the same profile neglecting the spin correlation effects, but
in practice the large integration error prevents this coincidence to
materialise in our simulation.

 Comparing figures \ref{figc} and \ref{figd} we see that rather than to
keep $p_T^{\rm cut}$ as a free parameter as one would do in an
experimental analysis, an easy way to simulate the effect of transverse
momenta cuts is to impose cuts at 100 GeV on both tau and missing
transverse momenta. The numerical integration for the background becomes
very unstable at this value and renders higher cuts impractical in our
simulation.

 The optimisation of cuts can be achieved by balancing between keeping the
production of the bosons $H^\pm$ and $W^\pm$ at low transverse momenta via
the cuts described above for example, and the final cuts on, or the
distribution of, the tau and missing transverse momenta. As we lower the
transverse momenta of the bosons we also lower the transverse momentum of
the top quark. Since we can not present a full simulation of top quark
identification near the beam directions we do not consider it worthwhile
speculating further into this, until we can utilise the full simulation.

 Table \ref{tabled} shows the result after the final cuts.

 The cut is very effective at large $M_{H^\pm}$, where it is clear that
larger values of $p_T^{\rm cut}$ would allow lower values of $\tan\beta$
to be probed. This can be easily achieved in practice by plotting the
cross section after cuts against $p_T^{\rm cut}$. The signal will show a
`bump' in the distribution, which can be confirmed as due to $H^\pm$
rather than statistical fluctuation by tau polarisation and spin
correlation analyses.

 For the intermediate values of $\tan\beta$ we can extrapolate the figures
by the known dependence of the cross section on $\tan\beta$. At
$M_{H^\pm}=200$ GeV, where the transverse momentum cut is least effective,
the signal-to-background ratio becomes 1:1 at $\tan\beta\sim2.3$. Given
the LHC integrated luminosity of 100 fb$^{-1}$ for the high luminosity
option, this corresponds to observing 3 events per year.

 For large values of $M_{H^\pm}$, say masses higher than 500 GeV, the
transverse momentum cut becomes very effective at reducing the background,
but we must deal with the exponentially falling cross section (see figure
\ref{figa}). As an example, at $M_{H^\pm}=1$ TeV, applying the same cuts
as above but setting $p_T^{\rm cut}=300$ GeV, we obtain a cross section of
0.00168 fb at $\tan\beta=1.5$. The numerical integration for the
background is very unstable, but we consider it negligible. We find that
in order to observe one event per year at the integrated luminosity of 100
fb$^{-1}$, $\tan\beta$ must be greater than 3.7.

We note that at such high values of $M_{H^\pm}$, the $bt$ decay mode
becomes relatively clean, as what we observe will be $b$ and $t$ jets
back-to-back with transverse momenta of about 500 GeV each. In the
$\tau\nu_\tau$ decay mode too, the signal will be spectacular, with the
hard tau and the missing momentum pointed back to back. Presumably the
requirement of top identification becomes less important here, and the
minimum $\tan\beta$ can be lowered significantly. As a conservative
estimate, $\tan\beta>3$ is accessible in the $\tau\nu_\tau$ decay mode of
$H^\pm$ for $M_{H^\pm}$ between 200 GeV and 1 TeV.

 For smaller values of $M_{H^\pm}$, the transverse momentum cut becomes
less effective, we must impose stricter preliminary cuts, and the analysis
becomes complex. On the other hand, this range involves kinematic
suppression on the $tb$ decay mode of $H^\pm$, and the branching ratio
into $\tau\nu_\tau$ becomes greater. As an added bonus, the $W^\pm h^0$
mode kicks in at low $\tan\beta$ where the $\tau\nu_\tau$ channel becomes
the least effective, which is easy to handle by imposing $b$ vertex
tagging on $h^0\to b\bar b$ since we will hopefully know the $h^0$ mass at
an early stage in LHC operation.

\section{Conclusions}

 We have discussed the search for the charged Higgs bosons $H^\pm$ of the
Two Higgs Doublet Model, inherent for example in \mssm, through the
$\tau\nu_\tau$ decay mode at LHC through the `$tb$ fusion' production
channel.

 We have found that after a combination of kinematic cuts, there is a vast
parameter space in which $H^\pm$ can be discovered, which have previously
been thought inaccessible at LHC.

 For $H^\pm$ masses heavier than 200 GeV, up to say 1 TeV, they can be
discovered for $\tan\beta\gtrsim3$. This is without the optimisation of
cuts.

 Provided we neglect the other decay modes which are model dependent, the
cross section is roughly proportional to $\tan^2\beta$, which implies that
if we improve the procedure merely by fourfold, which is not unrealistic,
the minimum $\tan\beta$ that can be probed falls to 1.5, which is below
the maximum $\tan\beta$ probed so far at LEP2 assuming \mssm. 

 The signal thus discovered can be confirmed by analysing the tau
polarisation and spin correlations.

 The plots of the tau and the missing
transverse momenta indicate that the $H^\pm$ mass can be evaluated to some
degree from the signal distribution provided there is large enough sample.
 At least before cuts, the $p_T$ distributions of the tau and the missing
transverse momenta are independent of $\tan\beta$, allowing one to
reconstruct $M_{H^\pm}$. The total cross section, being proportional to
$\tan^2\beta$, is then a good measure of $\tan\beta$. The spin
correlations involving the top quark decay products can be used to confirm
this measurement.

 The model dependent region near and below the $tb$ decay threshold, at
$M_{H^\pm}<200$ GeV, has not been treated in detail. A full study of this
region is desired, assuming \mssm, or otherwise.

 We only considered the irreducible background to this process. The
reducible background includes top pair production and $W^\pm+$jets
production. The second one is presumably removable by imposing the top
selection and as long as we impose some jet profile criterion in the
selection (so as to filter out soft multi-jets), it is at higher order in
\as\ anyhow. The top pair background is only significant if one of the
bottom quarks from top decay escapes in the beampipe. Although the cross
section for this may not be negligible, this is hardly likely to lead to
large tau and missing transverse momenta, as is the $W+$jets background,
and should be negligible in our analysis.

 We did not consider the supersymmetric background, notably the $2\to2$
sbottom--neutralino production from $gb$ fusion, with the sbottom decaying
into top quark, $W^\pm$ and neutralino. In some regions of the \susy\
parameter space where the sbottom decays into a top quark and a chargino
and the chargino decays into a $W^\pm$ and a neutralino, such background
will be large.

 Once $H^\pm$ has been discovered in this channel, the mass and the
$\tan\beta$ thus obtained can be used to optimise the signal extraction
procedure for other modes, notably the dominant $tb$ decay, leading to a
better understanding of the two-doublet Higgs sector.

 Our analysis is sensitive to top quark identification near the beam
direction, especially in the hadronic channel. This needs to be studied
with full jet and detector simulations, leading to the optimisation of the
signal extraction procedure, before the exact discovery contour can be
elucidated. We note that as long as top quark is identified as such, we do
not require much more information, other than the guarantee that there is
no other source of missing transverse momentum, for our analysis. In fact,
this guarantee may be sufficient on its own when using our procedure of
selecting high transverse momentum decay products from charged bosons
produced at low transverse momentum. The effect on reducible background of
relaxing the top quark identification requirement may thus be worth
analysing.

\subsection*{Acknowledgements}

 I thank Stefano Moretti for advice, discussions and for reading the
manuscript, and Bryan Webber for his guidance through my Ph.D.\ at
Cambridge.


\goodbreak


\clearpage\pagestyle{empty}

\subsection*{Table captions}

 \begin{table}[ht]
 \caption{Signal rates for some typical values of $M_{H^\pm}$ and
$\tan\beta$, and the `isotropic' and `scalar' background components.}
\label{tablea}
 \caption{Signal and background rates in the hadronic decay mode of the
top quark, after the selection cuts, before the final transverse momentum
cut.} \label{tableb}
 \caption{Signal and background rates in the hadronic decay mode of the
top quark, after the selection cuts, before the final transverse momentum
cut.} \label{tablec}
 \caption{Signal and background rates in the hadronic decay mode of the
top quark, after the final transverse momentum cut, with $p_T^{\rm
cut}=100$ GeV.} \label{tabled}
 \end{table}

\clearpage

\subsection*{Figure captions}

\begin{figure}[ht]
 \caption{Lowest order Feynman graphs for process (\ref{gbth}). Top and
$H^-$ decays are omitted for simplicity.} \label{feyn}
 \caption{(a -- top) The tree level $\tau\nu$ branching ratio of $H^\pm$
in the large mass limit, $M^2_{H^\pm}\gg m^2_t$, as functions of
$\tan\beta$.  (b -- bottom) The overall tree level $\tan\beta$ dependence
of production and $\tau\nu$ decay, in the large mass limit of $H^\pm$.}
\label{brplot}
 \caption{The total cross section in picobarns at LHC for $gb\to tH^-$, as
a function of $M_{H^\pm}$. $bg\to tH^-$ and the charge conjugate processes
are not included. Cross section must be multiplied by 4, the $\tan\beta$
dependence of (\ref{brnorm}) and the kinematic $bt$ suppression of
(\ref{exactbratios}).} \label{figa}
 \caption{Rapidity distributions for signal and background at LHC energy. 
$bg$ initial state is not included. Rapidity is positive along the
direction of the $g$ initial state parton.  Normalisation is to unity. 
Lines were used instead of histograms for the sake of visibility. Binning
width is 1. (a -- top) Signal events at $M_{H^\pm}=200$ GeV. (b -- middle)
Signal events at $M_{H^\pm}=500$ GeV.  (c -- bottom) Background.}
\label{figb}
 \caption{Transverse momentum distributions for signal and background at
LHC energy. Normalisation is to unity.  Lines were used instead of
histograms for the sake of visibility. Binning width is 50 GeV.  (a --
top)  Signal events at $M_{H^\pm}=200$ GeV. (b -- middle) Signal events at
$M_{H^\pm}=500$ GeV. (c -- bottom) Background.} \label{figc}
 \caption{Background transverse momentum distribution after the selection
cuts, before the final transverse momentum cut. Normalisation is to unity. 
See text for the cuts.} \label{figd}
 \end{figure}

\clearpage

 \begin{table}[p]\begin{center}\begin{tabular}{|c|c|c|}\hline
 $M_{H^\pm}$ (GeV) & $\tan\beta$       & $\sigma$ (fb)         \\\hline
 \multicolumn{3}{|c|}{$bg\to tH^-\to t\tau\bar\nu_\tau$} \\\hline
  200              &  1.5              & 4.78                  \\
  200              &  30               & 1530                   \\
  500              &  1.5              & 0.139                 \\
  500              &  30               & 52.1                  \\\hline
 \multicolumn{3}{|c|}{$bg\to tW^-\to t\tau\bar\nu_\tau$}       \\\hline
 \multicolumn{2}{|c|}{total}           &  13260                \\
 \multicolumn{2}{|c|}{`isotropic'}     &  6650                 \\
 \multicolumn{2}{|c|}{`scalar'}        &  6610                 \\\hline
 \end{tabular}\\\vspace*{1cm}Table 1\end{center}\end{table}

\clearpage

 \begin{table}[p]\begin{center}\begin{tabular}{|c|c|c|}\hline
 $M_{H^\pm}$ (GeV) & $\tan\beta$       & $\sigma$ (fb)         \\\hline
 \multicolumn{3}{|c|}{$bg\to tH^-\to t\tau\bar\nu_\tau$} \\\hline
  200              &  1.5              & 0.252                 \\
  200              &  30               & 81.0                  \\
  500              &  1.5              & 0.0205                \\
  500              &  30               & 7.69                  \\\hline
 \multicolumn{3}{|c|}{$bg\to tW^-\to t\tau\bar\nu_\tau$}       \\\hline
 \multicolumn{2}{|c|}{        }        & 0.86                  \\\hline
 \end{tabular}\\\vspace*{1cm}Table 2\end{center}\end{table}

\clearpage

 \begin{table}[p]\begin{center}\begin{tabular}{|c|c|c|}\hline
 $M_{H^\pm}$ (GeV) & $\tan\beta$       & $\sigma$ (fb)         \\\hline
 \multicolumn{3}{|c|}{$bg\to tH^-\to t\tau\bar\nu_\tau$} \\\hline
  200              &  1.5              & 0.388                 \\
  200              &  30               & 124                   \\
  500              &  1.5              & 0.0240                \\
  500              &  30               & 9.00                  \\\hline
 \multicolumn{3}{|c|}{$bg\to tW^-\to t\tau\bar\nu_\tau$}       \\\hline
 \multicolumn{2}{|c|}{        }        & 371                   \\\hline
 \end{tabular}\\\vspace*{1cm}Table 3\end{center}\end{table}

\clearpage

 \begin{table}[p]\begin{center}\begin{tabular}{|c|c|c|}\hline
 $M_{H^\pm}$ (GeV) & $\tan\beta$       & $\sigma$ (fb)         \\\hline
 \multicolumn{3}{|c|}{$bg\to tH^-\to t\tau\bar\nu_\tau$} \\\hline
  200              &  1.5              & 0.0123                \\
  200              &  30               & 3.94                  \\
  500              &  1.5              & 0.0177                \\
  500              &  30               & 6.659                 \\\hline
 \multicolumn{3}{|c|}{$bg\to tW^-\to t\tau\bar\nu_\tau$}       \\\hline
 \multicolumn{2}{|c|}{        }        & $0.030\pm0.009$       \\\hline
 \end{tabular}\\\vspace*{1cm}Table 4\end{center}\end{table}

\clearpage

 \begin{figure}[p]
 \centerline{\epsfig{figure=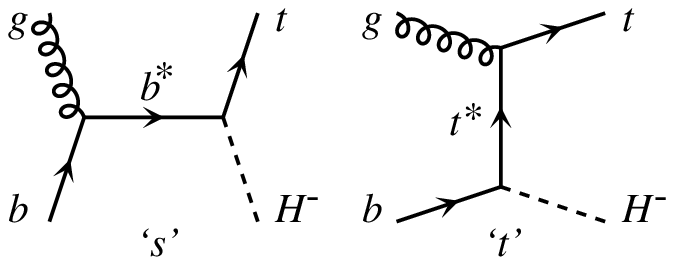,
           width=12cm,bbllx=0pt,bblly=0pt,bburx=200pt,bbury=100pt}}
 \centerline{Figure 1}
 \end{figure}

\clearpage

 \begin{figure}[p]
 \centerline{\epsfig{figure=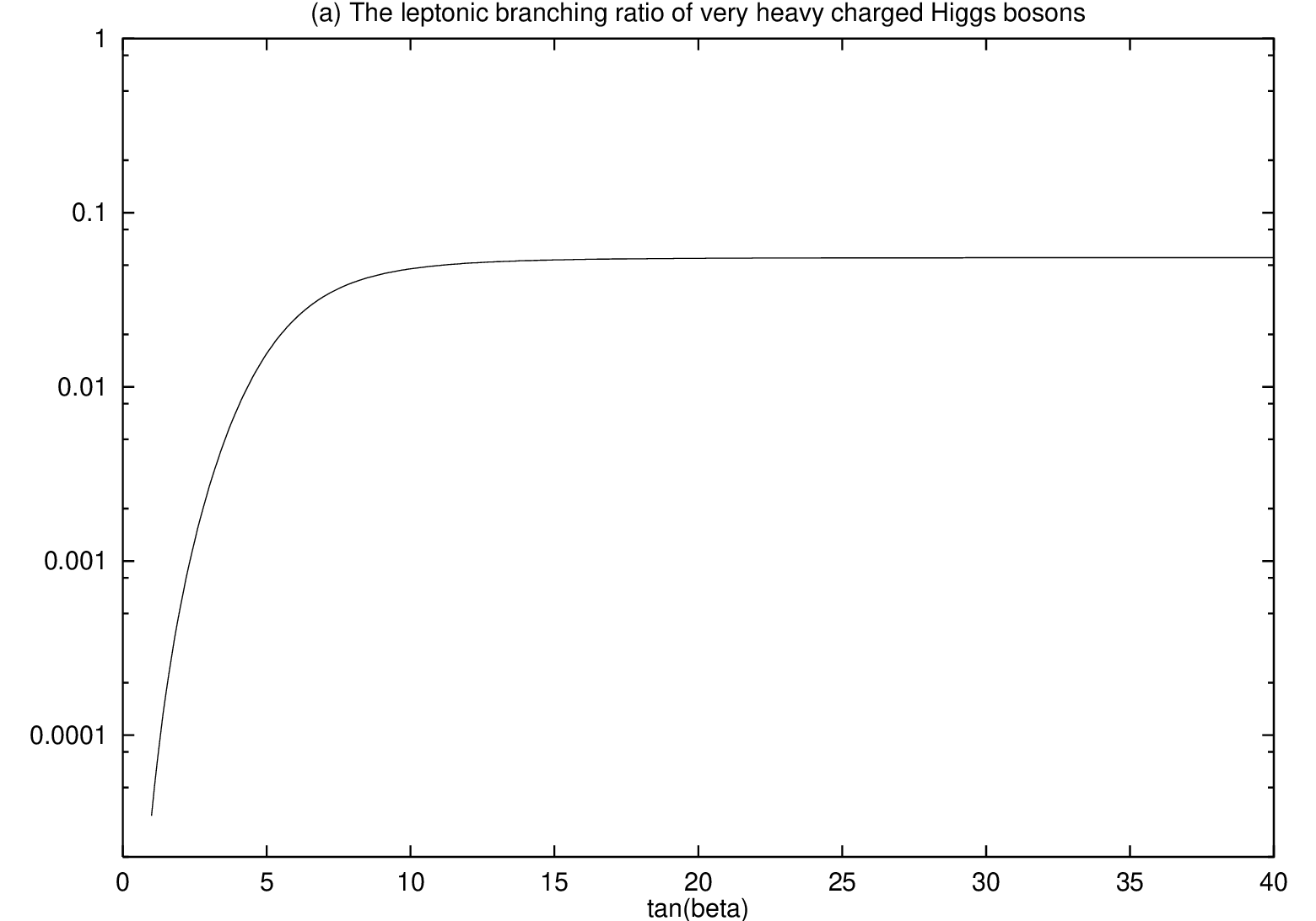,width=14cm,height=10cm}}
 \vspace*{0.5cm}
 \centerline{\epsfig{figure=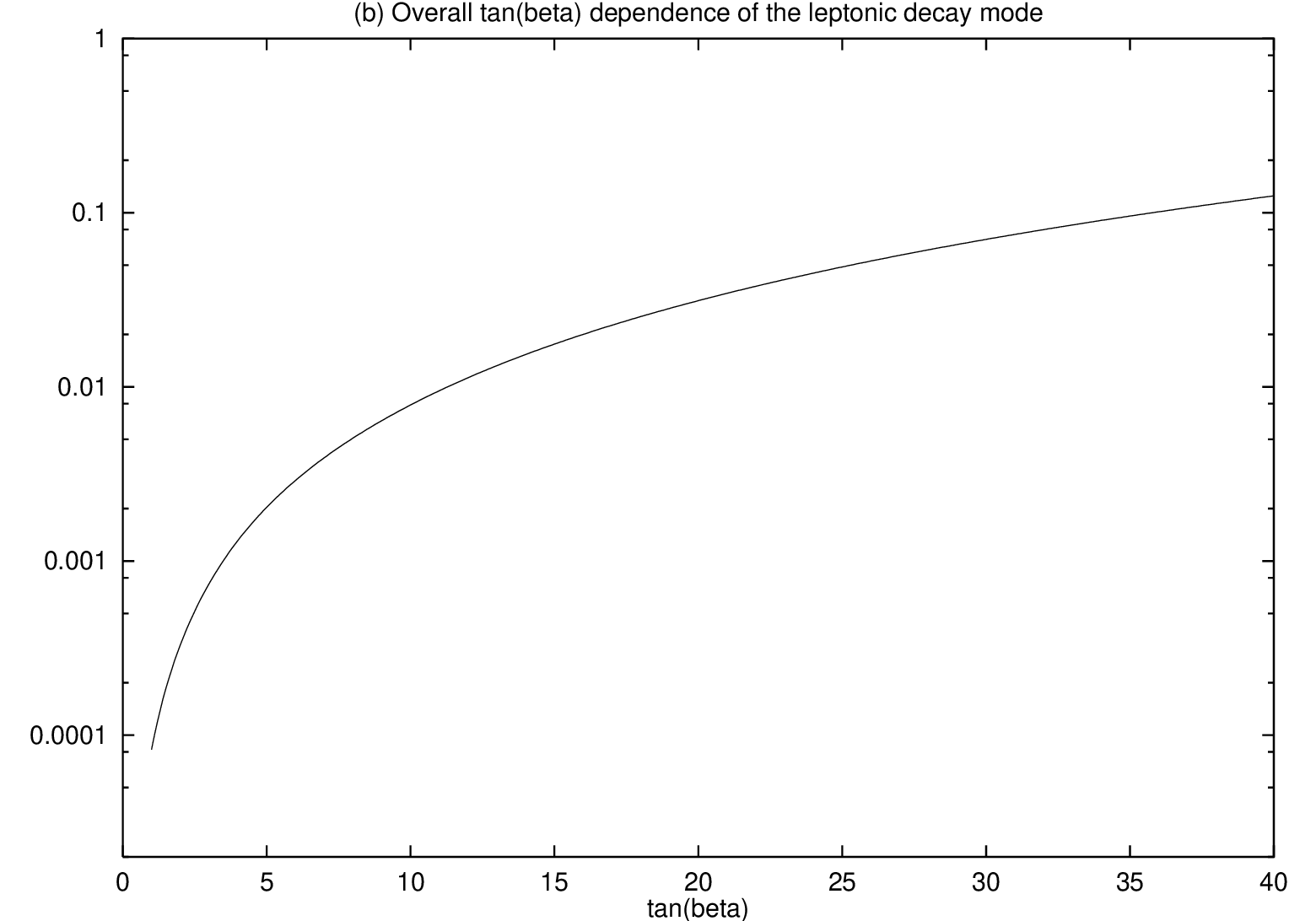,width=14cm,height=10cm}}
 \vspace*{0.5cm}
 \centerline{Figure 2}
 \end{figure}

\clearpage

 \begin{figure}[p]
 \centerline{\epsfig{figure=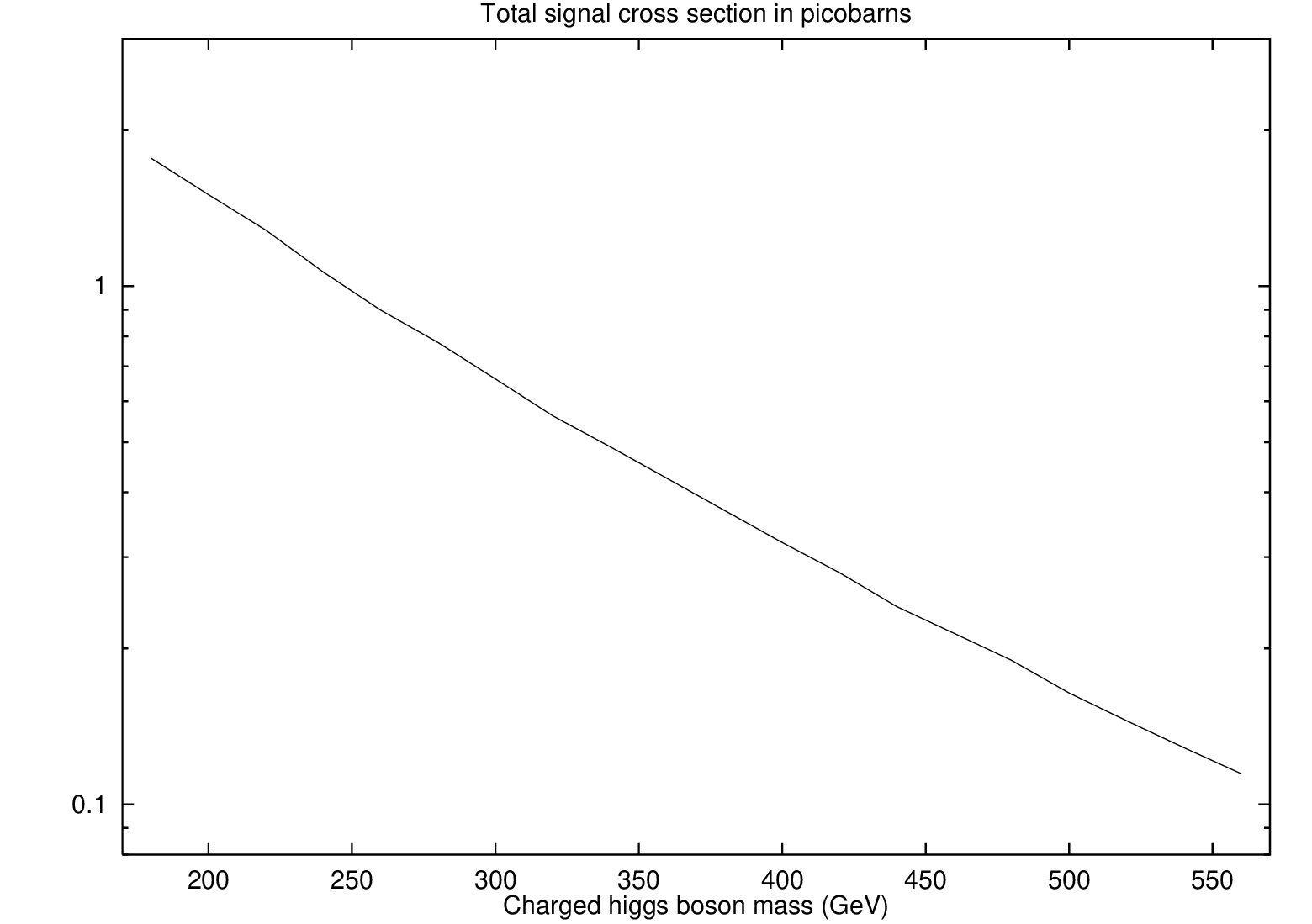,width=14cm,height=10cm}}
 \vspace*{0.5cm}
 \centerline{Figure 3}
 \end{figure}

\clearpage

 \begin{figure}[p] 
 \centerline{\epsfig{figure=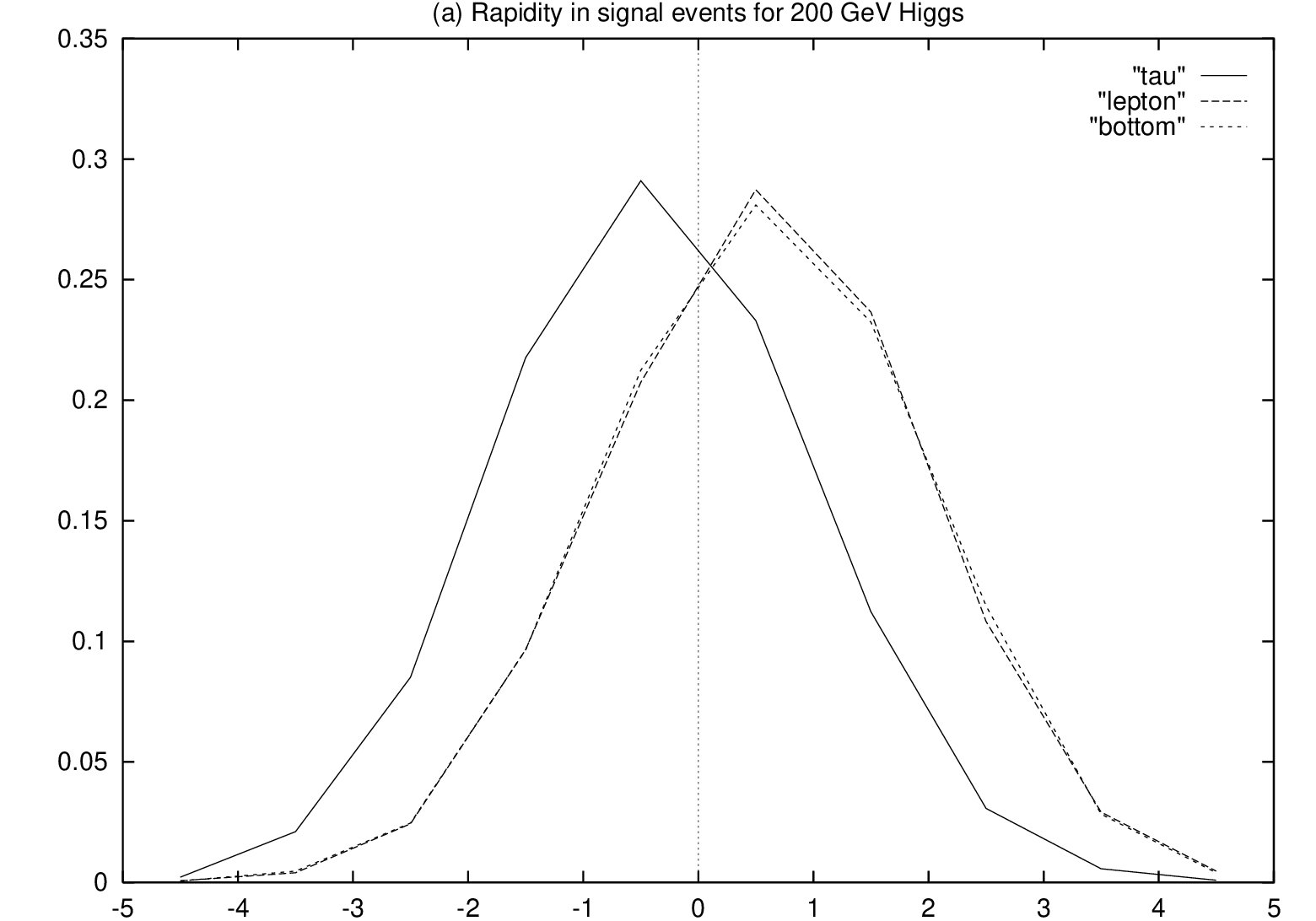,width=11cm,height=7cm}}
 \vspace*{0.5cm}
 \centerline{\epsfig{figure=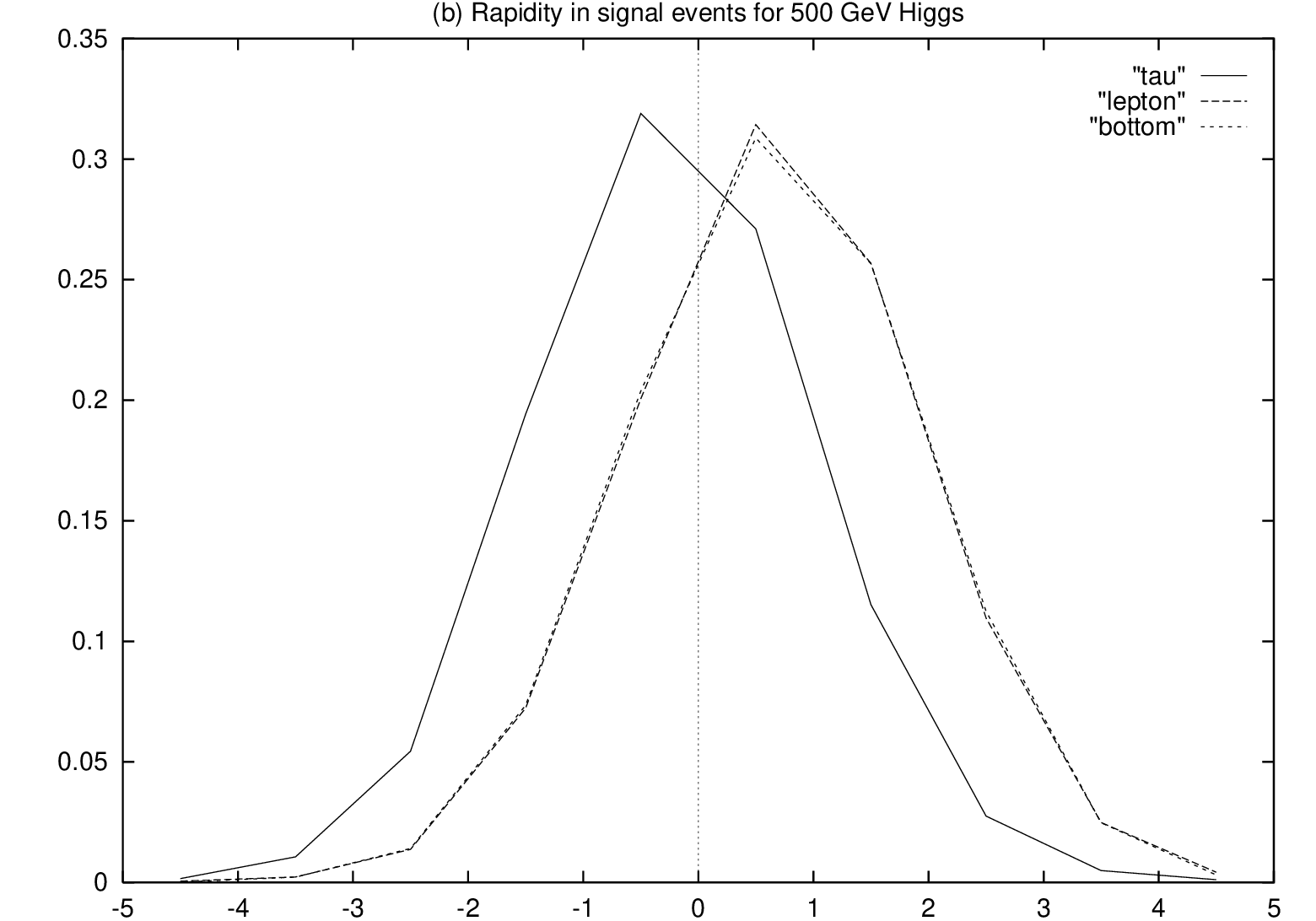,width=11cm,height=7cm}}
 \vspace*{0.5cm}
 \centerline{\epsfig{figure=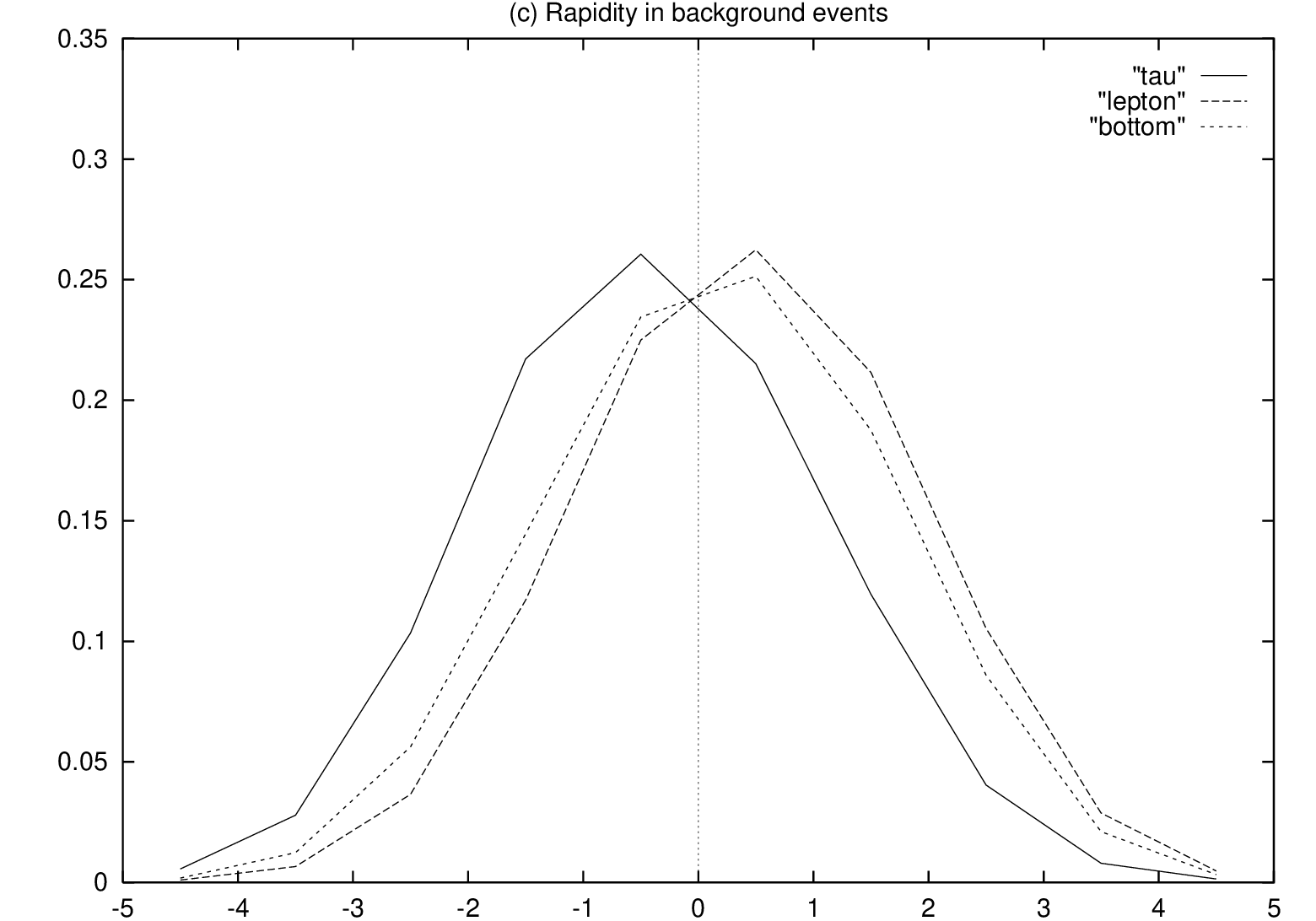,width=11cm,height=7cm}}
 \vspace*{0.5cm}
 \centerline{Figure 4}
 \end{figure}

\clearpage

 \begin{figure}[p] 
 \centerline{\epsfig{figure=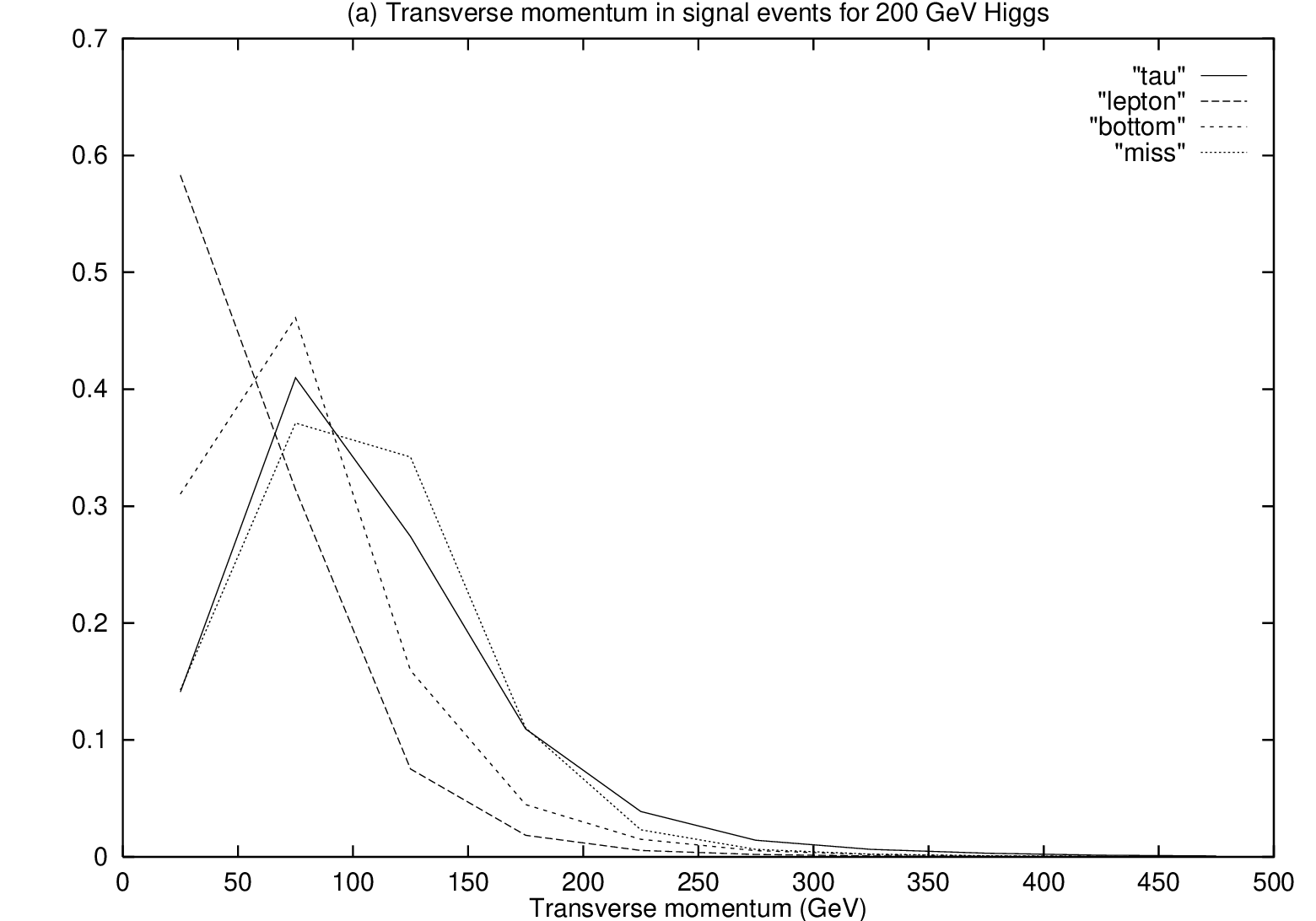,width=11cm,height=7cm}}
 \vspace*{0.5cm}
 \centerline{\epsfig{figure=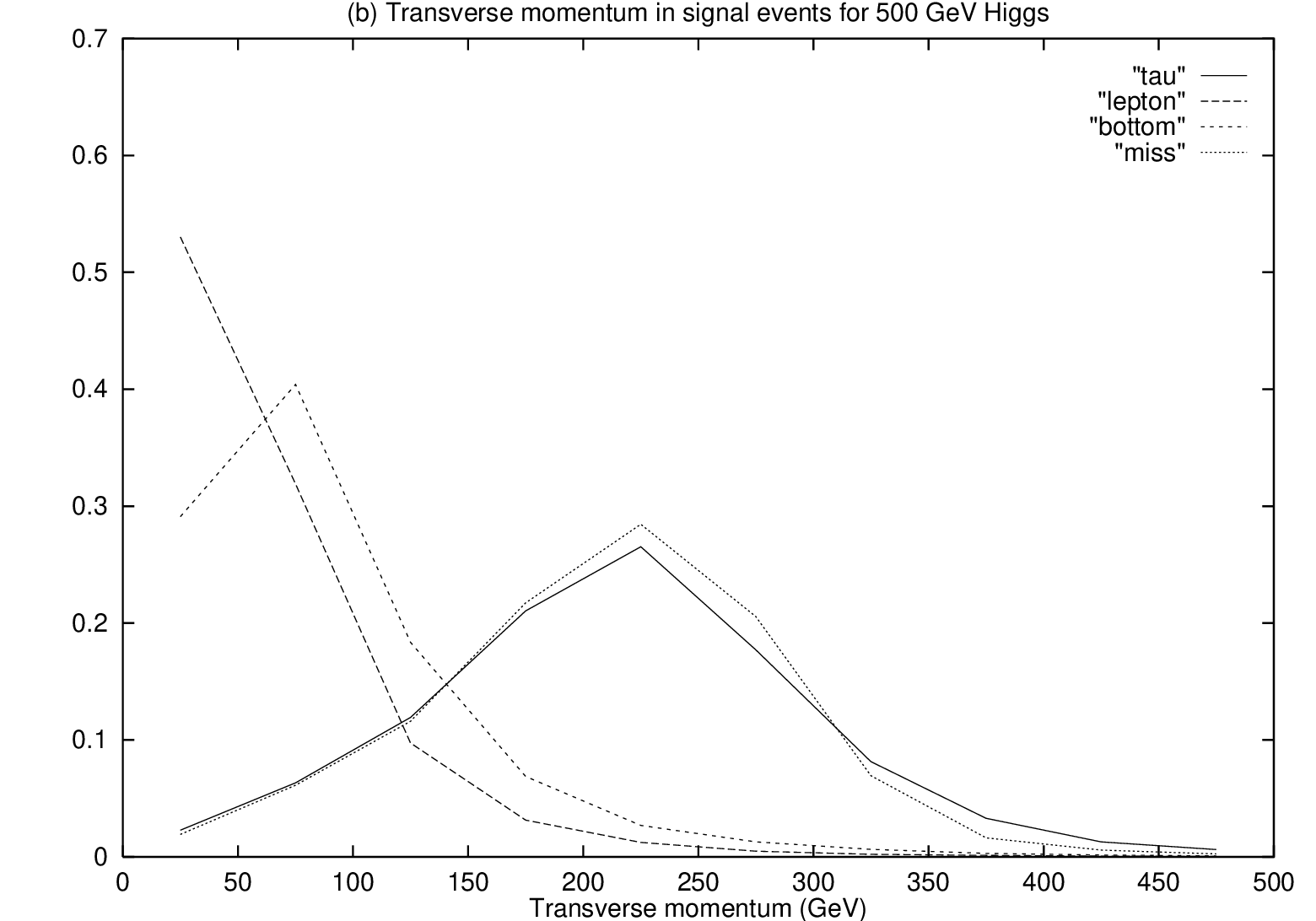,width=11cm,height=7cm}}
 \vspace*{0.5cm}
 \centerline{\epsfig{figure=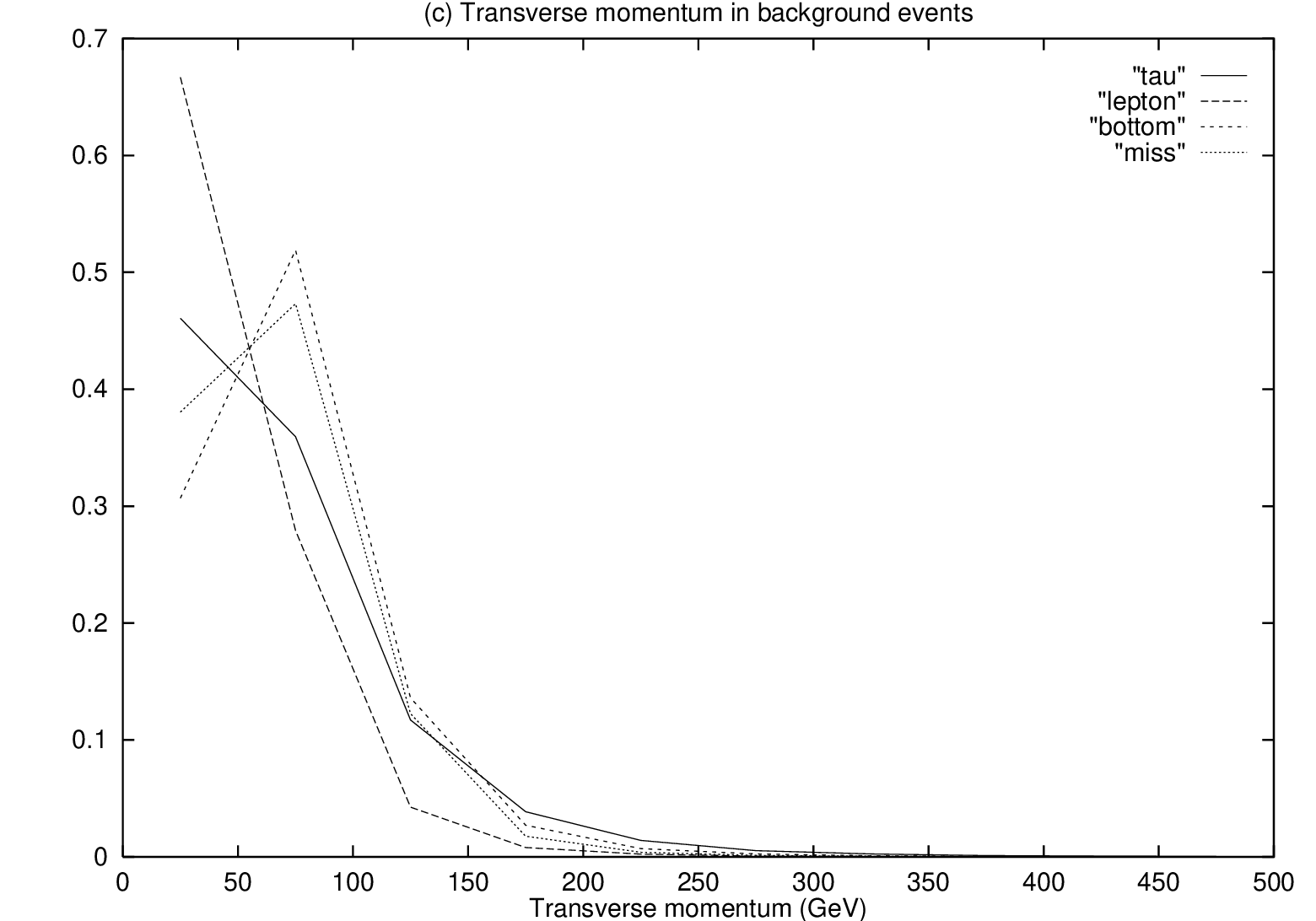,width=11cm,height=7cm}}
 \vspace*{0.5cm}
 \centerline{Figure 5}
 \end{figure}

\clearpage

 \begin{figure}[p] 
 \centerline{\epsfig{figure=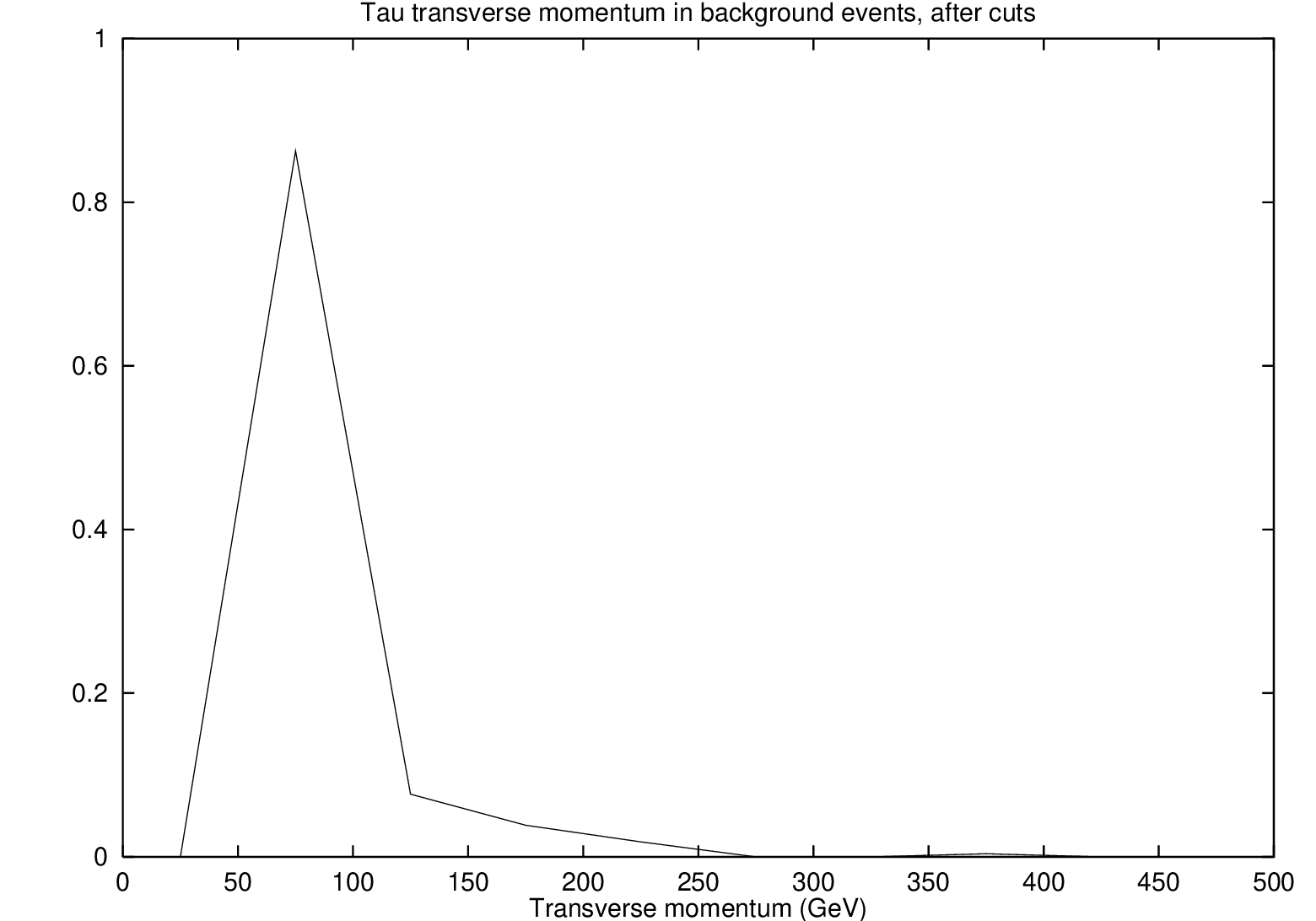,width=14cm,height=10cm}}
 \vspace*{0.5cm}
 \centerline{Figure 6}
 \end{figure}

\end{document}